\documentclass[12pt]{article}
\newcommand\unit[1]{\;{\rm #1}}
\newcommand\pd[2]{\frac{\partial #1}{\partial #2}}
\def\mref#1{(\ref{#1})}
\def\blambda{\lambda\!\!\!^{-}\,}
\begin{document}
\thispagestyle{empty} \vspace*{3cm}
    \begin{center}
 {\Large\sc $\pi$-mesons as the quanta of non-Newtonian hadronic fluid}

\bigskip
\bigskip
\bigskip
{\large Miroslaw Kozlowski and Janina Marciak-Kozlowska}

\bigskip
\bigskip
{\large Institute of Electron Technology\\
Al. Lotnik\'ow 32/46, 02-668 Warsaw, Poland\\e-mail:
MiroslawKozlowski@aster.pl}.
\end{center}

\vspace{2cm}
\begin{abstract}
It occurs that when we attempt to melt the nucleon in order to
obtain the free quark gas, or QGP fluid the mass of the heat
quantum (\textit{heaton}) is equal to the $\pi-$meson mass.

\end{abstract}

\newpage
\hspace{7cm}\parbox{8cm}{}\textit{\parbox{5cm}{Lasciate ogni speranza\\
{\small Dante Aligheri (1265--1321)}}}
\section{Introduction}
As is well known Nelson~\cite{1} in~1966 succeed in deriving the
Schr\"{o}dinger equation from the assumption that quantum
particles follow continous trajectories in a chaotic background.
The derivation of the usual linear Schr\"odinger equation follows
only if the diffusion coefficient~$D$, associated with quantum
brownian motion takes the value~$D=\hbar/2m$ as assumed by Nelson.

In the paper we study the transfer process of the quantum
particles in the context of the thermal energy transport in highly
excited matter. It will be shown that when matter is excited with
short thermal perturbation the response of the matter can be well
described by quantum hyperbolic heat transfer equation~(QHT) which
is the generalization of the parabolic quantum heat transport
equation~(PHT) with the diffusion coefficient \mbox{$D=\hbar/m$,}
where~$m$ denotes the mass of the diffused particles. The
obtained~QHT has the form
    \begin{equation}
    \frac1{c^2}\;\pd{^2T}{t^2}+\frac1{\tau c^2}\;\pd
    Tt=\frac{\alpha_i^2}3\,\nabla^2T,
    \end{equation}
where~$T$ denotes temperature, $c$~is the velocity of light
and~$\alpha_i=(1/137,0.15,1)$ is the fine structure constant for
electromagnetic interaction, strong interaction and strong
quark-quark interaction respectively, $\tau_i$  is the relaxation
time for scattering process.

When the QHT is applied to the study of the thermal excitation of the matter,
the quanta of thermal energy, the heaton can be defined with energies
$E_h^e=9\unit{eV}$, $E_h^H=7\unit{MeV}$ and $E_h^q=139\unit{MeV}$ for atomic,
nucleon and quark level respectively.

\section{The quantum heat transport equation}
One of the best models in mathematical physics is Fourier's model for the heat
conduction in matter. Despite the excellent agreement obtained between theory
and experiment, the Fourier model contains several inconsistent implications.
The most important is that the model implies an infinite speed of propagation
for heat. Cattaneo~\cite{2} was the first to propose a remedy. He formulated
new hyperbolic heat diffusion equation for propagation of the heat waves with
finite velocity.

There is an impressive amount of literature on hyperbolic heat
transport in matter~\cite{3,4,5}. In our book~\cite{6} we
developed the new hyperbolic heat transport equation which
generalizes the Fourier heat transport equation for the rapid
thermal processes. The hyperbolic heat conduction equation~(HHC)
for the fermionic system can be written in the form:
    \begin{equation}
    \frac1{\left(\frac13v_F^2\right)}\;\pd{^2T}{t^2}+
    \frac1{\tau\left(\frac13v_F^2\right)}\;\pd Tt=\nabla^2T,\label{eq1}
    \end{equation}
where $T$~denotes the temperature, $\tau$ -- the relaxation time
for the thermal disturbance of the fermionic system and $v_F$ is
the Fermi velocity.

In the subsequent we develop the new formulation of the HHC
considering the details of the two fermionic systems: electron gas
in metals and nucleon gas.

For the electron gas in metals the Fermi energy has the form:
    \begin{equation}
    E_F^e=(3\pi)^2\,\frac{n^{2/3}\hbar^2}{2m_e},\label{eq2}
    \end{equation}
where $n$ -- density and $m_e$ -- electron mass. Considering that
    \begin{equation}
    n^{-\frac13}\sim a_B\sim\frac{\hbar^2}{me^2},\label{eq3}
    \end{equation}
and $a_B$ -- Bohr radius, one obtains
    \begin{equation}
    E_F^e\sim\frac{n^{\frac23}\hbar^2}{m_e}\sim
    \frac{\hbar^2}{ma^2}\sim\alpha^2m_ec^2,\label{eq4}
    \end{equation}
where $c$ -- light velocity and $\alpha=1/137$ is the fine
structure constant. For the Fermi momentum, $p_F$ we have
    \begin{equation}
    p_F^e\sim\frac{\hbar}{a_B}\sim\alpha m_ec\label{eq5}
    \end{equation}
and for Fermi velocity,~$v_F$
    \begin{equation}
    v_F^e\sim\frac{p_F}{m_e}\sim\alpha c.\label{eq6}
    \end{equation}
Considering formula~\mref{eq6} equation~\mref{eq1} can be written as
    \begin{equation}
    \frac1{c^2}\;\pd{^2T}{t^2}+\frac1{c^2\tau}\;\pd Tt=
    \frac{\alpha^2}3\,\nabla^2T.\label{eq7}
    \end{equation}
As it is seen from~\mref{eq7} the~HHC equation is the relativistic
equation as it takes into account the finite velocity of light. In
order to derive the Fourier law from equation~\mref{eq7} we are
forced to break the special theory of relativity and put in
equation~\mref{eq7} $c\to\infty$, $\tau\to0$. In addition it was
demonstrated from~HHC in a~natural way, that in electron gas the
heat propagation velocity $v_h\sim v_F$ in the accordance with the
results of the pump probe experiments~

For the nucleon gas, Fermi energy is equal:
    \begin{equation}
    E_F^N=\frac{(9\pi)^{\frac23}\hbar^2}{8mr_0^2},\label{eq8}
    \end{equation}
where $m$ -- nucleon mass and $r_0$, which describes the range of
strong interaction is equal:
    \begin{equation}
    r_0=\frac\hbar{m_\pi c},\label{eq9}
    \end{equation}
$m_\pi$ -- is the pion mass. Considering formula~\mref{eq9} one
obtains for the nucleon Fermi energy
    \begin{equation}
    E_F^N\sim\left(\frac{m_\pi}m\right)^2mc^2.\label{eq10}
    \end{equation}
In the analogy to the equation~\mref{eq4} formula~\mref{eq10} can be written as
follows
    \begin{equation}
    E_F^N\sim\alpha_s^2mc^2,\label{eq11}
    \end{equation}
where $\alpha_s=0.15$ is fine structure constant for strong interactions.
Analogously we obtain for the nucleon Fermi momentum
    \begin{equation}
    p_F^N\sim\frac\hbar{r_0}\sim\alpha_smc,\label{eq12}
    \end{equation}
and for nucleon Fermi velocity
    \begin{equation}
    v_F^N=\frac{pF}m\sim\alpha_sc,\label{eq13}
    \end{equation}
and HHC for nucleon gas can be written as follows:
    \begin{equation}
    \frac1{c^2}\;\pd{^2T}{t^2}+\frac1{c^2\tau}\;\pd Tt=
    \frac{\alpha_s^2}3\,\nabla^2 T.\label{eq14}
    \end{equation}
In the following the procedure for the discretization of temperature~$T(\vec
r,t)$ in hot fermion gas will be developed. First of all we introduce the
reduced de~Broglie wavelength
    \begin{eqnarray}
    \blambda_B^e=\frac{\hbar}{m_ev_h^e}, &\qquad& v_h^e=\frac1{\sqrt3}\,\alpha
    c,\\ \nonumber
    \lambda_B^N=\frac{\hbar}{mv_h^N}, &\qquad& v_h^N=\frac1{\sqrt3}\,\alpha_sc,\label{eq15}
    \end{eqnarray}
and mean free path, $\lambda^e$, $\lambda^N$:
    \begin{eqnarray}
    \lambda^e&=&v_h^e\tau^e,\\ \nonumber
    \lambda^N&=&v_h^N\tau^N.\label{eq16}
    \end{eqnarray}
Considering formulae~\mref{eq15}, \mref{eq16} we obtain HHC for electron and
nucleon gases:
    \begin{eqnarray}
    \frac{\lambda_B^e}{v_h^e}\;\pd{^2T^e}{t^2}+\frac{\lambda_B^e}{\lambda^e}\;
    \pd Tt&=&\frac\hbar{m_e}\,\nabla^2T^e,\label{eq17}\\
    \frac{\lambda_B^N}{v_h^N}\;\pd{^2T^N}{t^2}+\frac{\lambda_B^N}{\lambda^N}\;
    \pd{T^N}t&=&\frac\hbar m\,\nabla^2T^N.\label{eq18}
    \end{eqnarray}
Equations~\mref{eq17} and~\mref{eq18} are the hyperbolic partial differential
equations which are the master equations for heat propagation in Fermi electron
and nucleon gases. In the following we will study the quantum limit of heat
transport in the fermionic systems. We define the quantum heat transport limit
as follows:
    \begin{equation}
    \lambda^e=\blambda_B^e,\qquad\lambda^N=\blambda_B^N.\label{eq19}
    \end{equation}
In that case equations~\mref{eq17}, \mref{eq18} have the form:
    \begin{eqnarray}
    \tau^e\,\pd{^2T^e}{t^2}+\pd{T^e}t&=&\frac\hbar{m_e}\,\nabla^2T^e,\label{eq20}\\
    \tau^N\,\pd{^2T^N}{t^2}+\pd{T^N}t&=&\frac\hbar m\,\nabla^2T^N,\label{eq21}
    \end{eqnarray}
where
    \begin{equation}
    \tau^e=\frac\hbar{m_e(v_h^e)^2},\qquad
    \tau^N=\frac\hbar{m(v_h^N)^2}.\label{eq22}
    \end{equation}
Equations~\mref{eq20}, \mref{eq21} define the master equation for quantum heat transport.
Having the relaxation time~$\tau^e$, $\tau^N$ one can define the
``pulsations''~$\omega_h^e$, $\omega_h^N$
    \begin{equation}
    \omega_h^e=(\tau^e)^{-1};\qquad\omega_h^N=(\tau^N)^{-1}\label{eq23}
    \end{equation}
or
    \[\omega_h^e=\frac{m_e(v_h^e)^2}\hbar,\qquad\omega_h^N=\frac{m(v_h^N)^2}\hbar\]
i.e.
    \begin{eqnarray}
    \omega_h^e\hbar&=&m_e(v_h^e)^2=\frac{m_e\alpha^2}3\,c^2\nonumber \\
    \omega_h^N\hbar&=&m(v_h^N)^2=\frac{m\alpha^2_s}3\,c^2.\label{eq24}
    \end{eqnarray}
Formulae~\mref{eq24} define the Planck-Einstein relation for heat
quanta~$E_h^e$, $E_h^N$
    \begin{eqnarray}
    E_h^e&=&\omega_h^e\hbar=m_e(v_h^e)^2\nonumber \\
    E_h^N&=&\omega_h^N=m_N(v_h^N)^2.\label{eq25}
    \end{eqnarray}
The heat quantum with energy~$E_h=\hbar\omega$ can be named as the
{\it heaton\/} in complete analogy to the {\it phonon\/}, {\it
magnon\/}, {\it roton\/}, etc. For $\tau^e$, $\tau^N\to0$
equations~\mref{eq20}, \mref{eq24} are the Fourier equations with
quantum diffusion coefficients~$D^e$,~$D^N$
    \begin{eqnarray}
    \pd{T^e}t&=&D^e\nabla^2T^e,\qquad D^e=\frac\hbar{m_e},\label{eq26}\\
    \pd{T^N}t&=&D^N\nabla^2T^N,\qquad D^N=\frac\hbar m.\label{eq27}
    \end{eqnarray}
The quantum diffusion coefficients $D^e$, $D^N$ for the first time
were introduced by E.~Nelson~\cite{1} and discussed in
papers~\cite{7,8,9}.

For finite $\tau^e$, $\tau^N$ for $\Delta t<\tau^e$, $\Delta t<\tau^N$
equations~\mref{eq20}, \mref{eq21} can be written as follows
    \begin{eqnarray}
    \frac1{(v_h^e)^2}\;\pd{^2T^e}{t^2}&=&\nabla^2T^e,\label{eq28}\\
    \frac1{(v_h^N)^2}\;\pd{^2T^N}{t^2}&=&\nabla^2T^N.\label{eq29}
    \end{eqnarray}

Equations~\mref{eq28}, \mref{eq29} are the wave equations for quantum heat
transport~(QHT). For $\Delta t>\tau$ one obtains the Fourier
equations~\mref{eq26}, \mref{eq27}.
\section{The possible interpretation of the \emph{heaton} energies}
First of all we consider the electron and nucleon gases. For electron gas we
obtain from formula~\mref{eq15}, \mref{eq25} for~$m_e=0.51\unit{MeV/{\mit
c}^2}$, $v_h=(1/\sqrt{3})\alpha c$
    \begin{equation}
    E_h^e=9\unit{eV},\label{eq36}
    \end{equation}
which is of the order of the Rydberg energy. For nucleon gases one obtains
($m=938\unit{MeV/{\mit c}^2}$, $\alpha_s=0.15$) from formulae~\mref{eq15},
\mref{eq25}
    \begin{equation}
    E_h^N\sim7\unit{MeV}\label{eq37}
    \end{equation}
i.e. average binding energy of the nucleon in the nucleus (``boiling''
temperature for the nucleus)

When the ordinary matter (on the atomic level) or nuclear matter (on the
nucleus level) is excited with short temperature pulses ($\Delta t\sim\tau$)
the response of the matter is discrete. The matter absorbs the thermal energy
in the form of the quanta~$E_h^e$ or~$E_h^N$.

It is quite natural to pursue the study of the thermal excitation to the
subnucleon level i.e. quark matter. In the following we generalize the QHT
equation~\mref{eq7} for quark gas in the form:
    \begin{equation}
    \frac1{c^2}\;\pd{^2T^q}{t^2}+\frac1{c^2\tau}\;\pd{T^q}t=\frac{(\alpha_s^q)^2}3
    \,\nabla^2T^q\label{eq38}
    \end{equation}
with~$\alpha_s^q$ -- the fine structure constant for strong
quark-quark interaction and~$v_h^q$ -- thermal velocity:
    \begin{equation}
    v_h^q=\frac1{\sqrt3}\,\alpha_s^qc.\label{eq39}
    \end{equation}
Analogously as for electron and nucleon gases we obtain for quark heaton
    \begin{equation}
    E_h^q=\frac{m_q}3(\alpha_s^q)^2c^2,\label{eq40}
    \end{equation}
where~$m_q$ denotes the mass of the average quark mass. For quark
gas the average quark mass can be calculated according to
formula~\cite{10}
    \begin{equation}
    m_q=\frac{m_u+m_d+m_s}3=\frac{350\unit{MeV}+350\unit{MeV}+550\unit{MeV}}3=
    417\unit{MeV},\label{eq41}
    \end{equation}
where~$m_u$, $m_d$, $m_s$ denotes the mass of the up, down and
strange quark respectively. For the calculation of
the~$\alpha_s^q$ we consider the decays of the baryon resonances.
For strong decay of the~$\Sigma^0(1385\unit{MeV})$ resonance:
    \[K^-+p\to\Sigma^0(1385\unit{MeV})\to\Lambda+\pi^0\]
the width~$\Gamma\sim36\unit{MeV}$ and lifetime~$\tau_s$
    \[\tau_s=\frac\hbar\Gamma\sim10^{-23}\unit{s}.\]
For electromagnetic decay
    \[\Sigma^0(1192\unit{MeV})\to\Lambda+\gamma,\]
$\tau_e\sim10^{-19}\unit{s}$. Considering that
    \[\left(\frac{\alpha_s^q}{\alpha}\right)\sim\left(\frac{\tau_e}{\tau_s}\right)
    ^{1/2}\sim100,\]
one obtains for~$\alpha_s^q$ the value
    \begin{equation}
    \alpha_s^q\sim1.\label{eq42}
    \end{equation}
Substituting formulae~\mref{eq41}, \mref{eq42} to formula~\mref{eq40} one
obtains:
    \begin{equation}
    E_h^q\sim139\unit{MeV}\sim m_\pi,\label{eq43}
    \end{equation}
where~$m_\pi$ denotes the $\pi$-meson mass. It occurs that when we
attempt to ``melt'' the nucleon in order to obtain the free quark
gas the energy of the \emph{heaton} is equal to the $\pi$-meson
mass (which consists of two quarks). It is the simple presentation
of quark confinement. Moreover it seems that the standard
approaches to the melting of the nucleons into quarks through the
heating processes in ``splashes'' of the chunks of the nuclear
matter do not promise the success.

\section{QGP as non-Newtonian fluid}
We argue that at excitation energy of the order of pion mass the
hadronic matter undergoes the phase transition to the
non-Newtonian fluid with the relaxation time, $\tau$
\begin{equation}
\tau=\frac{\hbar}{m_qc^2},\label{eq44}
\end{equation}
where $m_qc^2$ is of the order of 400~MeV and the velocity of
sound $v_h\sim c$. The thermal energy quanta in non-Newtonian
hadronic fluid is of the order of $140~{\rm MeV}\simeq m_{\pi}$.
In this model the $\pi-$meson can be described as the ``phonon''-
excitations of the non-Newtonian hadronic fluid. With the
excitation energy higher then 140~MeV the boiled fluid evaporates
the $\pi-$meson copiously. The spectra of the emitted $\pi-$mesons
can be described by formula~\cite{6}
\begin{equation}
d\eta=\frac{N_0mT^{-1}}{K_2\left(\frac{m}{T}\right)}c^{-3}\gamma^5u^2
\exp\left[-m\gamma T^{-1}\right]dV du.\label{eq45}
\end{equation}
In formula~(\ref{eq45})
$\gamma=\left(1-\frac{u^2}{c^2}\right)^{-\frac12}$, $T$ is the
temperature of the fluid and $dV$ is the volume element. Function
$K_2\left(\frac{m}{T}\right)$ is the modified Bessel function of
the second kind.

\end{document}